# I-V analysis of high-energy lithium-ion-irradiated Si and GaAs solar cells

*A. Meulenberg, Jr.,*[a] *B. Jayashree,*[b] *Ramani,*[b] *M. C. Radhakrishna,*[b] *and Saif Ahmad Khan*[c]

[a] Department of Instrumentation, Indian Institute of Science, Bangalore 560012, India
[b] Dept. of Physics, Bangalore University, Jnana Bharathi campus, Bangalore-560056 India
[c] Inter-university Accelerator Centre, Arun Asaf Ali Marg, New Delhi-110067, India,

**ABSTRACT:**

Space-grade Si and GaAs solar cells were irradiated with 15 and 40 MeV lithium ions. Dark-IV analysis (with and without illumination) reveals differences in the effects of such irradiation on the different cell types

## I. INTRODUCTION

Our prior work[1,2] has reported the results of high-energy lithium-ions onto Si and GaAs solar cells. Different damage modes for cells from the two materials could be seen from cursory examination of their dark I-V curves. The primary damage modes for the two cell types were attributed more to the cell structure (difference in active thickness and base doping) than to the material difference. A more detailed I-V analysis[3] is undertaken here to determine if further information could be extracted from the dark I-V curves measured under AM0 and zero-illumination levels.

Previous dark-IV analysis of electron- and proton-damaged Si and GaAs solar cells[4,5] was used as a basis for comparison with the present results of Li-ion irradiated cells. While most Si cells can be accurately modeled by a simple model consisting of the bulk (n=1), junction (n=2), and resistance ($R_s$ and $R_{sh}$) contributions (eq. 1, with constant values for $I_{sc}$, $I_{o1}$, and $I_{o2}$), the fit to the present GaAs cells is not as good and has been shown[6] to depend on additional parameters (with $I_{sc}$, $I_{o1}$, and $I_{o2}$ being functions of V)

$$I = I_{sc} - I_{01}[\exp\{q(V+IR_s)/kT\}-1] - I_{02}[\exp\{q(V+IR_s)/2kT\}-1] - (V+IR_s)/R_{sh} \qquad 1)$$

The same thing has been observed in the cells (unirradiated and Li irradiated) in this experiment. As a consequence, equation 1 is used to fit the Si data; but, an older/simpler form (eq. 2) is used for the GaAs cells. While the number of independent variables is not altered ($I_{o1}$ and $I_{o2}$ become $I_o$ and n) and the fit to data is improved by using equation 2, the ability to identify spatial and functional sources of dark current is reduced.

$$I = I_{sc} - I_0[\exp\{q(V+IR_s)/nkT\}-1] - (V+IR_s)/R_{sh} \qquad 2)$$

The physical basis for the distinction between the two equations is related to the ability to separate the bulk currents from the junction currents. In the silicon cells, there is a clear distinction between the high-field junction region and the quasi-neutral bulk regions (even when the strong doping gradient of a shallow-junction emitter region is included). There are two reasons why this separation is less possible for the GaAs cells: the probable graded-doping of the bulk region and the AlGaAs "window" on the front of the cell.

Graded doping of the GaAs bulk region has two benefits. First, the high built-in voltage of the GaAs creates very strong fields in the junction region. These fields would normally be accentuated by the reduced junction-region thickness resulting from the heavily-doped base needed to maximize the cell Voc. These strong fields increase the bandgap narrowing in this region (from the Franz-Keldysh effect)[3], which would increase the junction recombination current.

## II. RESULTS : DARK IV ANALYSIS - - SILICON

Fig. 1 displays the dark-IV analysis for one of the silicon cells before irradiation. Fig. 1a is for the unilluminated case and Fig. 1b is for the AM0 case. Note the excellent fit to data using eq. 1. Such good fits provide a sensitive measure of the shunt resistance (9 kΩ +/- 10%) and the dominant dark junction-recombination current contribution ($I_{jr}$ = 48 nA +/- 10%) for Fig. 1a. The series resistance and diffusion current interfere with each other in this calculation and therefore are not as reliably determined without going to higher currents, which requires pulsed measurements to prevent cell heating during the measurement. On the other hand, in Fig. 1b, the bulk diffusion current dominates and therefore, the series resistance can be accurately determined.

There are strong differences in the AM0 and unilluminated dark IV curves. The obvious noise level ($\sigma$ = ~0.2%) at low voltages, resulting from fluctuations in the solar simulator, limits the sensitivity in the determination of $R_{sh}$ in Fig. 1b. The dramatic increase in the dark-diffusion current ($I_d$, the n=1 component) with illumination is particularly important in terms of the AM0 electrical characteristics. Illumination also appears to increase the junction-leakage current (lowers $R_{sh}$). However, with the low-voltage noise levels, this observation may not be as real (even though there is a logical mechanism for it).

Fig. 2 provides the same information for a silicon cell that has been exposed to $5 \times 10^{11}$ Li ions/cm$^2$. Again, eq. 1 fits the data very well. The greatest effect of radiation is seen in the dark-diffusion currents. Comparison of Figs. 1a and 2a shows the largest change in $I_d$ (0.05 pA up to 340 pA). All other changes are within about a factor of 3.

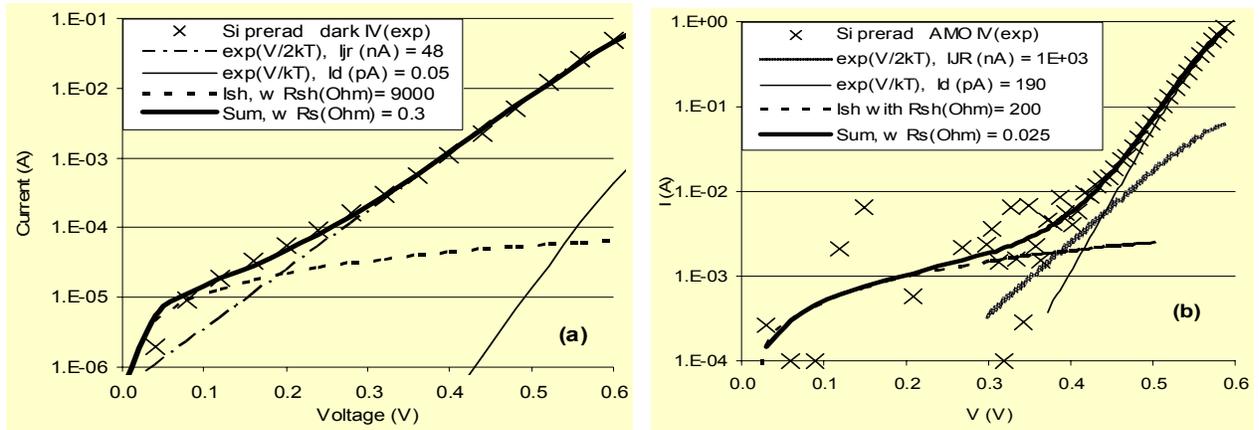

*Fig 1. Dark IV characteristics for an a) unilluminated and b) AM0 illuminated silicon solar cell. Contributing dark-current components and resistances for the unirradiated cell are identified and quantified.*

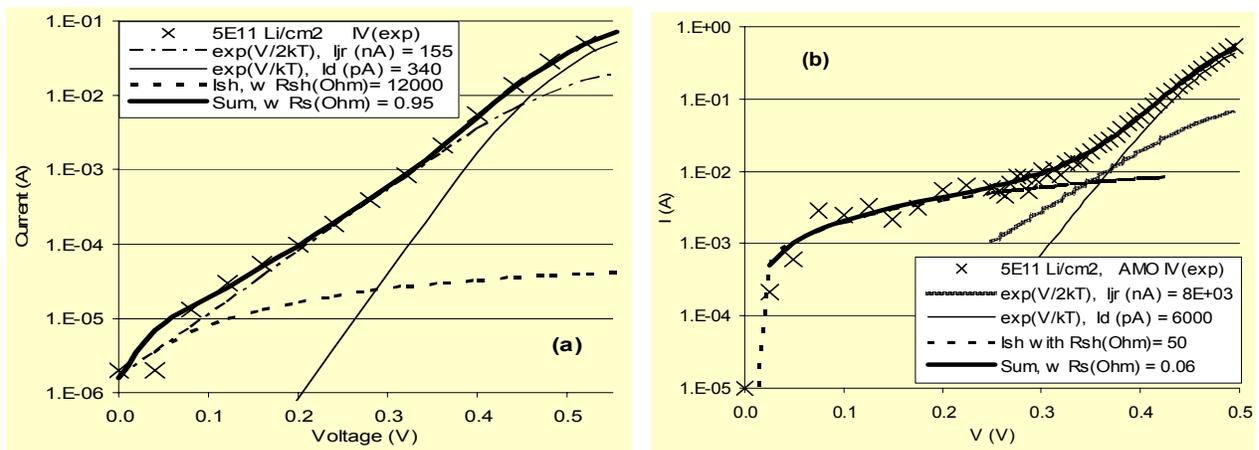

*Fig 2. Dark IV characteristics for the silicon solar cell in Fig.1, both a) unilluminated and b) AM0 illuminated, but after irradiation with $5 \times 10^{11}$ /cm$^2$ 40 MeV Li ions.*

Comparison of Figs. 1b and 2b also shows the largest change to be in $I_d$. However, in this case the change is much less because the starting value is much higher (190 pA vs 0.05 pA). Nevertheless, the final value is higher than that in the dark (6000 pA vs 340 pA). The net change in $I_d$ with irradiation is from 190 pA to 6000 pA. Since the high $I_d$ in Fig. 1b is related to the photo-carrier density, the 6000 pA value would likely be higher were it not for the reduced $I_{sc}$ resulting from the reduced minority-carrier lifetime. The primary cause of this change in $I_{sc}$ is the introduction of defects into the bulk of the cell, which lowers the minority-carrier lifetime. This reduces the diffusion length of the cell which in turn eliminates the benefits of the $p^+$ back contact to both the $I_{sc}$ and the $V_{oc}$. The change in series resistance (~ 3x) is comparable to that observed in the proton-irradiated GaAs cells of Reference 6 and may also be related to the reduction in photo-carrier density resulting from the reduced lifetime.

### III. RESULTS : DARK IV ANALYSIS - - GAAS

Despite the need to use eq. 2, rather than eq. 1, dark I-V analysis for GaAs cells (irradiated with 15 and 40 MeV Li ions, Figs. 3-6), indicates similar trends as those for the silicon cells. Illumination increases the dark current (particularly for unirradiated cells), and decreases the bulk-series and the junction-shunt resistance. Li irradiation greatly increases the dark current in the cells (much more so in the illuminated case for GaAs than for Si) and lowers the junction-shunt resistance. The major difference in the Si and GaAs radiation trends is that the high-n component of dark current increases most with fluence in the GaAs case and the low-n component of dark current increases most with fluence in the Si case.

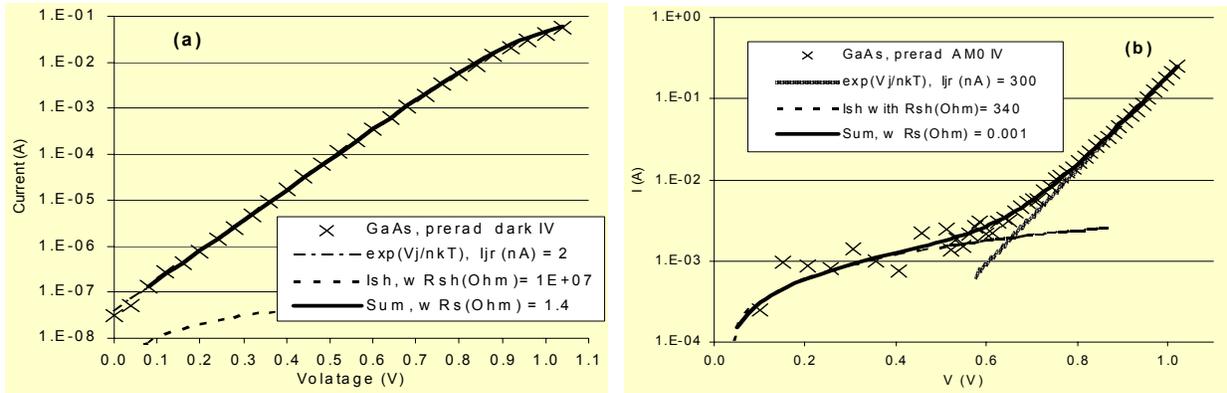

*Fig 3. Dark IV characteristics for an a) unilluminated (n = 2.52) and b) AM0 illuminated (n = 2.88) GaAs/Ge solar cell prior to irradiation.*

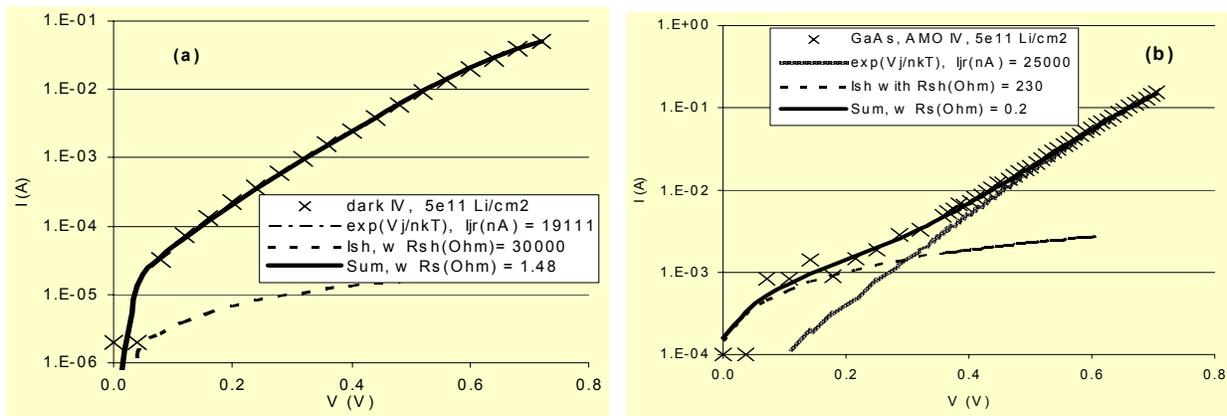

*Fig 4. Dark IV characteristics for the GaAs/Ge solar cell in Fig 3, both a) unilluminated (n = 3.14) and b) AM0 illuminated (n = 3.14), but after irradiation with $5 \times 10^{11}$ /cm$^2$ 40 MeV Li ions*

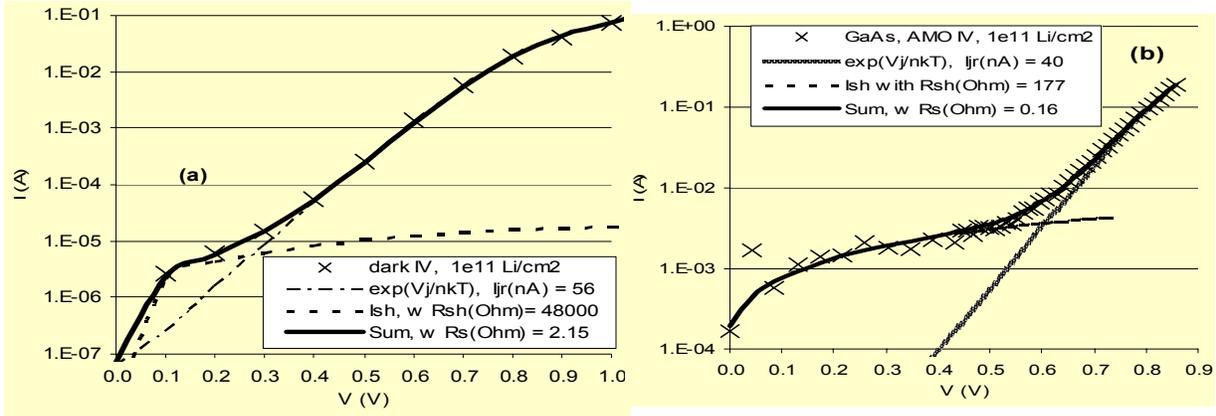

Fig 5. Dark IV characteristics for a GaAs/Ge solar cell, both a) unilluminated (n = 2.3) and b) AM0 illuminated (n = 2.16), after irradiation with $1\times10^{11}$ /cm$^2$ 15 MeV Li ions

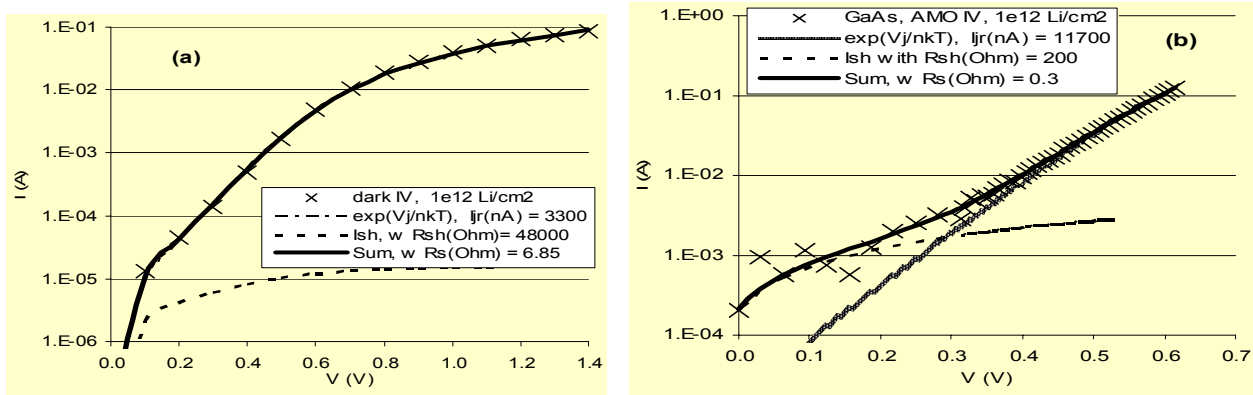

Fig 6. Dark IV characteristics for a GaAs/Ge solar cell, both a) unilluminated (n = 3) and b) AM0 illuminated (n = 2.58), after irradiation with $1\times10^{12}$ /cm$^2$ 15 MeV Li ions

Table I provides representative examples of fitted values for unirradiated and irradiated, unilluminated and AM0, Si and GaAs solar cells. Observed differences are as describe for the figures, but quantified. The causes are explored and explained[3,7] in terms of cell structure, the displacement damage, and the type of defects from 15 and 40 MeV Li ions to Si and GaAs solar cells.

TABLE I   Constant values in characteristic equations for Si and GaAs cells

|  | Fluence (#/cm$^2$) | Ijr(nA) | Id(pA) | Io | n | Rs | Rsh |
|---|---|---|---|---|---|---|---|
| **Si** | Unirradiated | 48 | 0.05 | - | - | 0.3 | 9k |
| Fig.1 | AM0 | 1000 | 190 |  |  | 0.025 | 200 |
| 40 MeV | 5E11 | 155 | 340 |  |  | 0.95 | 12k |
| Fig.2 | AM0 | 8000 | 6000 |  |  | 0.06 | 50 |
| **GaAs** | Unirradiated | - | - | 2 | 2.52 | 1.4 | 10M |
| Fig.3 | AM0 |  |  | 300 | 2.88 | < 0.01 | 340 |
| 40 MeV | 5E11 |  |  | 19000 | 3.14 | 1.48 | 30k |
| Fig.4 | AM0 |  |  | 25000 | 3.14 | 0.2 | 230 |
| **GaAs** | 1E11 (15 MeV) | - | - | 56 | 2.3 | 2.15 | 48k |
| Fig.5 | AM0 |  |  | 40 | 2.16 | 0.16 | 177 |
| 15 MeV | 1E12 |  |  | 3300 | 3 | 6.85 | 48k |
| Fig.6 | AM0 |  |  | 11700 | 2.58 | 0.3 | 200 |

Use of non-ionizing-energy loss (NIEL) values for predicting radiation damage from different sources may be influenced by the observed dependence of cell characteristics on illumination (minority-carrier, high-injection effects). Most solar cell comparisons are made with parameters measured under illumination. Nevertheless, even for equal displacement damage energy from a particle, the effects on lattice damage now depends not only on the dominant defects created, but also on the injection level. This dependence of different defects on injection level means that I-V analysis can provide information that would help identify such differences and would aid in evaluating the efficacy of simulations of and with different radiation sources.

## IV. CONCLUSIONS

The use of Li ions, rather than protons, emphasizes the cluster defect damage over the more dominant point defect damage of the lighter particles. The differences in structure of the Si cells (dark current dominated by the bulk contribution) vs the GaAs cells (junction recombination dominated) accentuate this difference further. It is expected that, if GaAs cells with different structure (doping levels, multiple junction, etc.) are exposed to this type radiation, then differences within the GaAs types will be amplified by irradiation with Li when compared to that from protons. Depending on the application, this can be useful or a problem.

## V. ACKNOWLEDGMENT

B. Jayashree is deeply indebted to the management, A.P.S Educational Trust, Bangalore and the University Grants Commission (UGC) New Delhi, India for allowing her to pursue research work in Bangalore University, India.
This work is supported in part by: HiPi Consulting, New Market, MD, USA; by the Science for Humanity Trust, Bangalore, 560094, India; and the Science for Humanity Trust, Inc, Tucker, GA, USA.